# Metabolism of Social System
## NIPD in RBN

Deni Khanafiah[1], Hokky Situngkir[2]

[1] Research assistant in Bandung Fe Institute, Indonesia, mail: <denig10@eudoramail.com>
[2] Departement Computational Sociology, Bandung Fe Institute, Indonesia, mail: hokky@ee.itb.ac.id, web: <http://www.geocities.com/quicchote>

## Abstract

Random Boolean Network has been used to find out regulation patterns of genes in organism. This approach is very interesting to use in a game such as N Person PD. Here we assume that action is influenced by input in the form of choices of cooperate or defect he accepted from other agent or group of agents in the system. Number of cooperators, pay off value received by each agent, and average value of the group pay off, are observed in every state, from initial state chosen until it reaches its state cycle attractor. In simulation performed here, we gain information that a system with large number agents based on action on input K equals to two, will reach equilibrium and stable condition over strategies taken out by its agents faster than higher input, that is K equals to three. Equilibrium reached in longer interval, yet it is stable over strategies carried out by agents.

## 1. Introduction

One of interesting aspects in understanding complex social system is about how simple interactions occur among elements constitutes it: human individual interactions *emerge* as complex macro phenomena (Paczuski, 2000). In Sociology, this concept later known as *human agency* (Fuch, 2002; Situngkir, 2003a) or structure of macro-micro linkage in social system (Sawyer, 2001; Situngkir, 2003b).

Macro-micro linkage of the social system stated that macro phenomenon in the form of social structure (norms, social behaviors, organization, etc) are emergent phenomena coming out from interactions occurring in individual/micro level (Situngkir, 2003a; Frumkin, *et al*). The idea is similar to some sociologists who support structuration theory such as Giddens (1984) and Minger who also see that social structure is a macro phenomena resulted from an individual level interactions process (Fuch, 2002).

In Game Theoretic Analysis, in this case Prisoner's Dilemma, interactions among individuals viewed as mathematic model where those interactions interpreted only through watching its action of choose to cooperate or not to cooperate (*defect*). Simple interaction occurred in Prisoner's Dilemma is believed to be a form of more elementary interaction than another more complicated intrapersonal interactions.

In analysis using Prisoner's Dilemma, every individual action is not based on certain social structure; namely social norms, social behavior/habit or even social sanction (Robinson et al, 2000), henceforth we can virtually see whether certain social structure such as cooperation property is likely to emerge as a collective character from selfish agents that play. Human agency concept stating that social structure may emerge through interactions among their agents is highly appropriate in case of Prisoner's Dilemma game. This game also has been known to be utilized in understanding complex phenomena occur in social and field of economy (Cho, 2000).

In social system as well as in nature, it is known that someone's action certainly affects the whole result gained by each person. *'Free rider'* occasionally appears as a *defector* who takes advantage from a group of cooperators. The case which Hardin (1968) later named as *Tragedy of Common* happened in many cases of natural resources exploitation, which are commonly were not encountered within internal economy. Over-exploitation of one natural resource certainly will cause the decreasing result or advantages that we can actually get.

As mentioned in the tragedy of common, social problems occur is hardly seen from individual level (microstructures). This is because each action chosen by an individual influenced by other individual actions in the system. To overcome this problem, evolutionary dynamic approach from *Prisoner's Dilemma* model should be one of the ways to see individual situation and strategy that results maximum yields (Lindgren, *et al*, 1998).

One of emergent phenomena from individual interactions in N-Person Iterated Prisoner's Dilemma/NIPD game is *cooperation* among agents. This phenomenon has been widely observed in various approaches such as *spatial models* (Scweitzer, 2002), *mean-field model* (Lindgren, 1998) and *Ring Model* (Alexander, 2002), and other forms of models. This cooperative phenomenon is interesting to observe, since it often happens in other systems, like company organization, living cells, up to molecular level (Axelrod, 1984).

In the rest of the paper, we will show how N-Person IPD applied into the Random Boolean Network model. This paper is divided into several sections; the first explains what N-Person Prisoner's Dilemma is and followed by some description about the Random Boolean Network. Thus will be explained how to use Random Boolean Network to describing N-Person Prisoner's Dilemma. This paper ends with some simulation results along with theoretical cause-effect performed.

## 2. N-person Iterated Prisoner's Dilemma

Prisoner's Dilemma is a game models interaction between two persons, where each individual or agent plays without knowing other agent's choice whether to cooperate (C) or not to cooperate/defect (D). Payoff that each agent will gain will depend on her choice and other agent's choice. Payoff value received by each agent is commonly expressed in form of table matrix as seen in Table 1.

**Table 1**
Pay-off matrix of 2-person Prisoner's Dilemma

|          | *cooperate* | *defect* |
|----------|-------------|----------|
| *cooperate* | R | S |
| *defect*    | T | P |

Where R represents the reward, T for temptation, S for sucker, and P for punishment.
Dilemmatic condition will rise as these below conditions acquired:
1. T>R>P>S
2. 2R>S+T

By noticing those pay-offs, in one round play, whatever action taken by other agent, each agent will tend to choose *cooperate* (C) – Nash equilibrium from this single play is easily understand that agent will tend to cooperate. However, if the game has a chance to play once more, then each agent will evaluate payoff gained and chosen in the previous round. Such game is called iterated Prisoner's Dilemma.

This IPD model is then developed into a game involving higher number of agents, be it for 3 persons (Akiyama, et al, 1995) or for N-person (Axelrod, 1997; Akimov, 1994; and Lindgren, 1998). This kind of game has been widely used to explain the emergence of cooperation in group of agents where every agent in the group is selfish and interacting playing Prisoner's Dilemma game.

In one round game, there will be *n* agents that choose 2 options of actions served – just as well as in 2-person PD, they are to cooperate (C) or to defect (D). Payoff achieved by agents will depend on number of cooperators, say it *p* numbers, so each agent will gain payoff for each option she takes, be it cooperate or defect. We may state payoff value received by agent as $\phi_C^p$ = pay-off value obtained by cooperate agent and $\phi_D^p$ = pay-off obtained by defect agent. Value $\phi_D^p$ should be higher than $\phi_C^p$, yet this pay-off value ($\phi$) for each agent will rise along with the rise of number person who chooses cooperate, and so the whole agents pay-off in the group. This dilemma can be formulated as follows:

1. $\phi_D^p > \phi_C^p$ …………………………..(1)
2. $\phi_D^{p+1} > \phi_D^p ; \phi_C^{p+1} > \phi_C^p$ ……………..(2)
3. $\phi_C^{p+1}(p+1) + \phi_D^{p+1}(N-(p+1)) > \phi_C^p p + \phi_D^p (N-p)$ ………………..(3)

In one round game of the NIPD, rational option that is most likely to be taken by each agent is *defect*, yet if all players choose defect, thus each player will surely gain lower value than if all choose to cooperate.

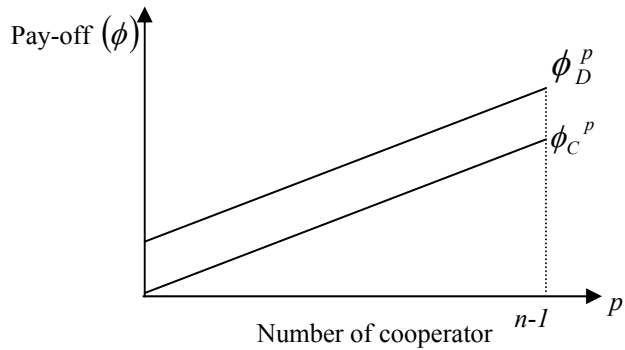

**Figure 1**
Graphic of pay-off value in n-IPD for each agent

Describing agent interactions in n-IPD has been improved widely and commonly there are two types of models widely used to describe interactions occur: *mean-field model* where all agents interact with all agents and *spatial model,* where agents lie in spatial distribution then interact locally. N-agents game becomes one choice of game model that is frequently used to understand aggregation from inter-component interaction to macro level in an evolutionary system (example: biology or social).

## 3. Random Boolean Network

Random Boolean Network (RBN) is generally constituted by certain numbers of node linked to input-output wire. For every time, each node states whether on or off as output, this output will become input to other nodes linked to it – and vice versa other node output will become input node for other nodes connected to it. Number of inputs linked to one node is equal to the connection with other nodes connected to it (possibly, plus one as the state from the node itself come to determine its next state). Further elaboration can be seen in Kauffman (1993).

Initial condition of on/off states is assigned with binary variables 0 and 1 and in the following time (**t+1**) will be updated to suits rule transition of the state it has. This rule transition is equal to Boolean Function where number of the function will equal to $2^{2^K}$. K represents the number of input. In other words, rule transition will describe nodes states at each time depend on the arrangement of input that flows.

RBN is basically a directed graph, *G = (N, E),* which can be represented with *N = (1,…, n)*, representing the number of nodes, and *K = (1,…,k),* representing the number of *input* for each *node* and rule transition $f_i = \langle f_1,...,f_n \rangle$, that it owns.

If a node, say *i,* has K neighbors *K =(1,…, k)*, and the state of neighbors (which becomes input for node *i*) is stated as $N_i = (i_1,...,i_k)$, thus we state $S_t^i$ as a condition of *node i* at time t, and $S_0^i$ stating initial condition of node *i*, then at each time the node will update its condition follows the following rule:

$$S_t^i = \begin{cases} f_i(S_{t-1}^{i_1},...,S_{t-1}^{i_k}) & if\ t>0; i_1,...i_k \in N_i \\ S_0^i & others \end{cases} \quad \text{…………………………..(4)}$$

RBN application was pioneered by Stuart A. Kauffman (1993) to show the self-organizing patterns in collective set of autocatalysis and genes regulation in the body of organism. Through RBN, the dynamic patterns of system constituted by many elements whose inter-regulating connection - as well as in gene regulation and autocatalysis set - becomes possible to learn. Kauffman observed dynamic behavior of RBN in collective set autocatalysis and gene regulation by varying N (the number of system components) and K (input of components). From those variations, he found that system behavior affected by K value. From K value variation, system behavior is said to compartmentalize into three regions, they are order region → complex → chaos (Kauffman, 1993:188-203). Division of the three regions is made based upon the dimensions describing the characteristic of the system, i.e.: state cycle, number of the state cycle attractor, and its adaptation over perturbation forming the structure transformation and minimum perturbation. The minimum perturbation is performed by changing the binary condition of one constituent's activity to its reverse activity while structural perturbation is performed by changing *Boolean* function from one function to a different function (Kauffmann, 1993:209-221).

Order region, that is for RBN with K=1, assigned with crystallizing pattern in the system, most of elements present in separated loop with derivatives of state chain – system does not show dynamical patterns. Complex area or sometimes-called "the edge of chaos" occurs at K~2. In this area system behavior assigned with long relatively small state cycle and quite stable to perturbations –will return to initial basin attractor when it is perturbed. While in chaos area - occur when value K>2, is characterized with long and huge state cycle, large amount number of state cycle and is very fragile over perturbation even only with minimum change. Furthermore, the behavior of the state cycle is hard to predict and depends on its initial condition.

In a chaotic system, region with value K>2, we also can have order region by adding big value of variable ***P*** or homogeneity parameter. Homogeneity parameter is fraction of 1 or 0 from output of Boolean function, where the value will range over 0.5 – 1.0. From observation carried out by Bernard and Gerard (Kauffmann, 1993), it is discovered that if a network added by increasing value ***P***, from 0.5 – 1.0, then behavior of a network would be getting to the order region. At ***P*** = 0.5, network becomes relatively chaotic while at ***P*** reaching 1.0 networks is relatively order. Hence, at certain network there will be found critical value of ***$P_c$*** that bridges chaos region and order one (Kauffman, 1993: 477-478).

## 4. Gene Regulation and NIPD

Looking at the requirements of the dilemma occur in NIPD represented by equation (1), (2), and (3), we can arrange a pay-off matrix for N-agents as seen in **Table 2**.

**Table 2**
Table pay-off matrix for n-agent Prisoner's Dilemma

|  | Number of Agent cooperates (cooperator) | | | | |
|---|---|---|---|---|---|
|  | $p_1$ | $p_2$ | $p_i$ | | $p_N$ |
| Cooperate (C) | $\phi_C^1$ | $\phi_C^2$ | —— | $\phi_C^i$ | —— | $\phi_C^N$ |
| Defect (D) | $\phi_D^1$ | $\phi_D^2$ | —— | $\phi_D^i$ | —— | —— |

Pay-off matrix in **Table 2** can be formulated as follow:

a. Pay-off for agent that chooses *cooperate*

$$\overline{\phi}_C = \frac{(R \; x \; p) + (S \; x \; (N-p))}{N} \quad \ldots\ldots\ldots\ldots\ldots\ldots\ldots\ldots\ldots\ldots\ldots(5)$$

b. Pay-off for agent chooses *defect*

$$\overline{\phi}_D = \frac{(T \; x \; p) + (P \; x \; (N-p))}{N} \quad \ldots\ldots\ldots\ldots\ldots\ldots\ldots\ldots\ldots\ldots(6)$$

If we illustrate agent behavior equals to a gene regulation process in metabolism of organism body, where those genes will come whether to a catalyzed state or inhibited depends on other agent activities that becomes input to it, thus the same tool to explain that kind of regulation, then RBN can be applied in N-person Prisoner's Dilemma game.

For an illustration, we will take example of 3-agents NIPD using RBN approach (**Figure 3**). For 3 agents interacting randomly as well as in RBN, there will be 8 possible states (**Table 3**) as initial state. Those states will flow into next state, according to rule transition (Boolean function) that each agent has.

**Table 3**
Possible *state* for 3 Boolean agents

| 1 | 2 | 3 |
|---|---|---|
| 0 | 0 | 0 |
| 0 | 0 | 1 |
| 0 | 1 | 0 |
| 0 | 1 | 1 |
| 1 | 0 | 0 |
| 1 | 0 | 1 |
| 1 | 1 | 0 |
| 1 | 1 | 1 |

We have previously known that with K inputs for each node there will be $2^{2^K}$ numbers of Boolean functions. In example of three agents in **Figure 3**, we see that input received by each agent is 2, so that there will be $2^{2^2} = 16$ possible Boolean functions (Red'ko, 1998). Those functions will illustrate agent condition at t+1, when certain input combination received at time t (**Table 4**).

**Table 4**
Possibility of Boolean Function for each agent

| Input at time (t) | | Agent 3 choice at t+1 | | | | | | | | | | | | | | |
|---|---|---|---|---|---|---|---|---|---|---|---|---|---|---|---|---|
| 1 | 2 | 1 | 2 | 3 | 4 | 5 | 6 | 7 | 8 | 9 | 10 | 11 | 12 | 13 | 14 | 15 | 16 |
| 0 | 0 | 0 | 0 | 0 | 0 | 1 | 0 | 0 | 1 | 0 | 1 | 1 | 0 | 1 | 1 | 1 | 1 |
| 0 | 1 | 0 | 0 | 0 | 1 | 0 | 0 | 1 | 1 | 1 | 0 | 0 | 1 | 0 | 1 | 1 | 1 |
| 1 | 0 | 0 | 0 | 1 | 0 | 0 | 1 | 1 | 0 | 0 | 0 | 1 | 1 | 1 | 0 | 1 | 1 |
| 1 | 1 | 0 | 1 | 0 | 0 | 0 | 1 | 0 | 0 | 1 | 1 | 0 | 1 | 1 | 1 | 0 | 1 |

If the above Boolean Function is indexed as $f_i$, where i = (1,2,…,16) as in **Table 4,** thus, there will be Boolean Function $f_i$ of 16 for 2 inputs case. If we define agent strategy as set of choices whether to cooperate or defect taken by agent, where choices taken will suit a transition rule that it has, thus $f_i = \langle f_1, f_2, ..., f_{16} \rangle$, will perform possible strategy that can be chosen by each agent in the system.

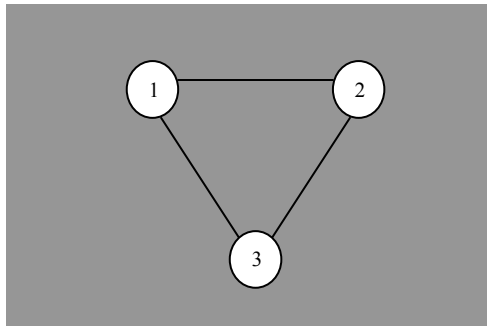

**Figure 3**
Interaction model for 3 agents in RBN

While each agent choose every strategy randomly from 16 existing function and one initial state is chosen, it will flow the next state according to the rule transitions until it reaches an attractor state. For example, we assume agent 1 choose strategy 2, agent 2 choose strategy 9 and agent 3 choose strategy 10, then we will obtain function as in **Table 5**.

**Table 5**
Example of Transition Rule for 3 agents Boolean

| t | | t+1 |
|---|---|---|
| 1 | 2 | 3 |
| 0 | 0 | 1 |
| 0 | 1 | 0 |
| 1 | 0 | 0 |
| 1 | 1 | 1 |

| t | | t+1 |
|---|---|---|
| 1 | 3 | 2 |
| 0 | 0 | 0 |
| 0 | 1 | 1 |
| 1 | 0 | 0 |
| 1 | 1 | 1 |

| t | | t+1 |
|---|---|---|
| 2 | 3 | 1 |
| 0 | 0 | 0 |
| 0 | 1 | 0 |
| 1 | 0 | 0 |
| 1 | 1 | 1 |

With transition rule/Boolean function chosen, we will have the set of state from the system at each time as seen in **Table 6**.

**Table 6**
State transition for 3 agents Boolean

| t | | | t+1 | | |
|---|---|---|---|---|---|
| 1 | 2 | 3 | 1 | 2 | 3 |
| 0 | 0 | 0 | 0 | 0 | 1 |
| 0 | 0 | 1 | 0 | 1 | 1 |
| 0 | 1 | 0 | 0 | 0 | 0 |
| 0 | 1 | 1 | 1 | 1 | 0 |
| 1 | 0 | 0 | 0 | 0 | 0 |
| 1 | 0 | 1 | 0 | 1 | 0 |
| 1 | 1 | 0 | 0 | 0 | 1 |
| 1 | 1 | 1 | 1 | 1 | 1 |

The change of state from time to time suits the transition rule that each agent has (Table 6) will form a certain set pattern that is a cyclic called state cycle (**Figure 4**).

By forming the set of state pattern, we can calculate number of aspects in Prisoner's Dilemma, i.e.: the number cooperative and defective agent, pay-off received by agent or pay-off gained by the group of agents at every state (t, t+1, t+2, …) from initial state up to the state of attractor.

Then we can define the loose agent as the agent with lower play-off and the winner whose highest pay-off. In the next round, the agent evaluates the strategy it has and will pick a better strategy from other possible Boolean Function. Here we have evolutionary NIPD, where one's strategy may adapt with other strategy until she gains most mutual condition.

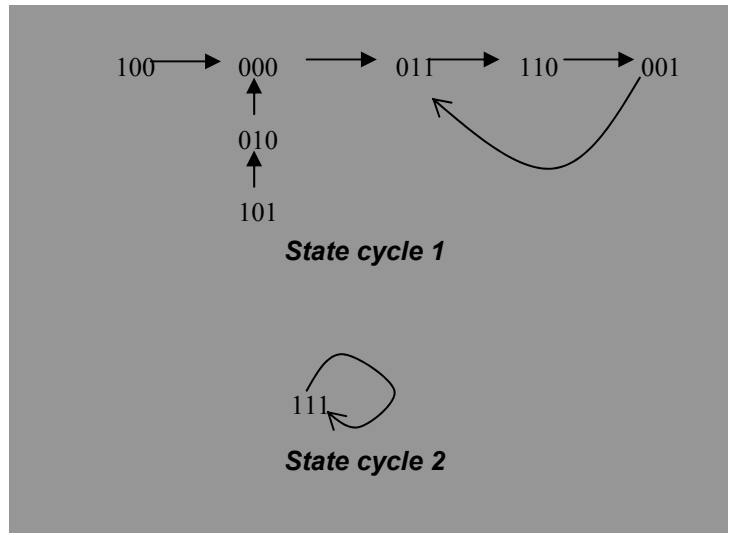

**Figure 4**
Example of state cycle pattern for 3 Boolean agents

Such 3 agents RBN model can be modified further to larger agents model with input varied from 1 to N. By flowing the whole pay-off from the system at initial state to the attractor state, and pay-off gained by every agent, we can observe which strategy that is most beneficial for each agent and beneficial to the group.

## 5. Model Inspiration

From the above explanation, we perform a set of experiment to see how IPD applied in RBN model. Before that, we have seen and analyzed another models that similar with our model i.e. Alexander (2002) and Paczuski et al (2000) models and compared it to the actual RBN model with some needed modifications on implementation to the RBN model so that it will really describe individual interactions as represented in IPD.

In the ring model by Alexander (2002), the translation of N-person Prisoner's Dilemma was performed in Random Boolean Network fashion. He assumed agents play Prisoner's Dilemma in a network with ring formation, where each agent updates its state by rule: imitating winner neighbor state in order each agent calculates how much pay-off gained by every of his neighbors. Every agent assumed to know other agent state that is other neighbor's neighbor. At that condition, agents playing in ring model described as a node that has input from its 2 closest neighbors node and the node itself (**Figure 5**). In this ring PD game, every agent supposed to link with 2 neighbors and play 2-person PD game. Pay-off obtained by agent only encountered only from neighbor agents, thus the social dilemma can not be described in such fashion.

Different with Alexander (2002), Paczuski, *et. al*. (2000) described agents as well as in Random Boolean Network – with a goal to describe the economic model, i.e.: minority game, in RBN fashion. They viewed that every agent will ground her action on limited information he received from other agent or other group of agent. This condition was described in NK Random Boolean Network, where N represents the number of agent and K represents the number of input of agents or other group of agent. Agent's strategy is made based on input she has, that picked from the possible Boolean Function. At every round every agent will gain reward and punishment. After certain interval, the looser, that is agent lies in majority group, will change her strategy randomly.

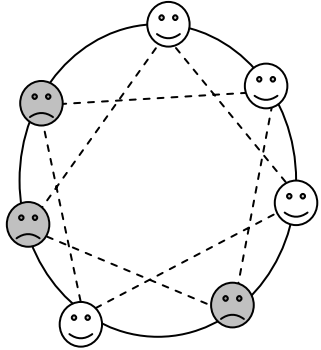

**Figure 5**
RBN for agents played in ring model of n-person Prisoner's Dilemma

Inspired by those models, we try to construct a new model of NIPD. In the model of Alexander (2002), we found that interaction among agents was only limited in the pre-defined ring, while human interactions model is highly vary and highly random interconnectedness. Many agents network that has inter-regulating connection is more proper to be described in NK Boolean Network model.

Actions chosen by agent are varied and cannot be simplified into one strategy only. There will be numbers of strategy chosen by an agent in a group. From this point, we can see strategy as action probability taken responding every set of actions taken by other agents (input). This kind of strategy can be represented with possible Boolean Function. Every agent will decide whether to cooperate or defect, and as the consequences, she will get certain pay-off as well as in NIPD. Each agent, from the initial states to the last state will get certain pay-off. Agent who gets lowest pay-off may change her strategy by imitating strategy chosen by the winner. Further, we can see the evolution of the system in its macro state.

## 6. Simulation Results

In this simulation, we use NIPD model in RBN with pay-off rule as expressed in equation (5) and (6). In this model choosing neighbor and initial strategy is performed randomly. Strategy of each agent chosen from Boolean rule table suits the K value used that is for K=2 used Boolean table of dimension 64x8[1]. Differ to model used by Alexander (2002), in this model every agent chooses strategy not option of cooperate or defect.

The simulation result for 20 agents and K=2 is described in **Figure 6**. This result showing that the dominant strategy at the end of simulation is the second strategy of Boolean rule: cooperative when all neighbor agents cooperative. In **Figure 7** we show a static (*frozen*) experimental result with the same initial setting.

From this point, we can say that with K=2 and homogeneity parameter for each epoch (long-time scale where each agent changes her strategy) as shown in the picture gaining frozen or periodic depends upon strategy in the initial condition of each agent. In **Figure 8,** we show the simulation result when agent number N=100. In this simulation the agents change their strategies in every 10-rounds - by assumption that length attractor of the state cycle is ($\sqrt{N}$) where **N** is number of agents. The changes are conducted by the rule that each agent imitate highest pay-off neighbor at every attractor length. **Figure 8** shows that as the number of cooperator increases, the average pay-off of the group increases too.

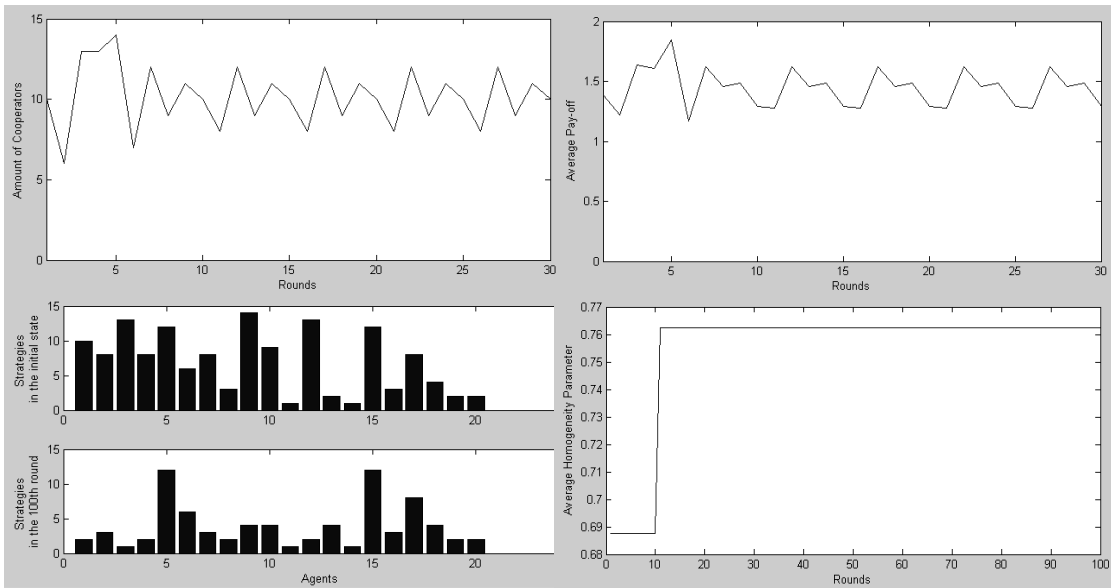

**Figure 6**
Simulation result with K=2, N=20, and 100 times iteration. Initial condition chosen randomly results stable aggregate condition and periodic at its equilibrium

---

[1] For value K>3, for example K=4 uses Boolean function matrix with large magnitude, that is 65536x16 – a very large dimension which is really difficult to run with computational resources that is used in this research.

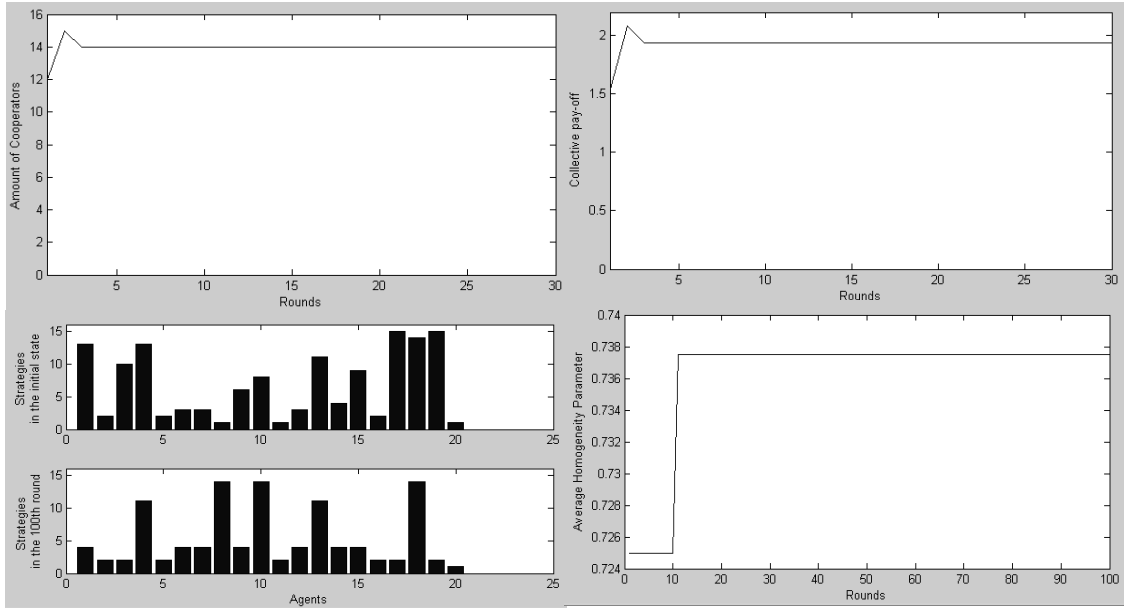

**Figure 7**
Simulation result for K=2 and N=20 with frozen result at equilibrium

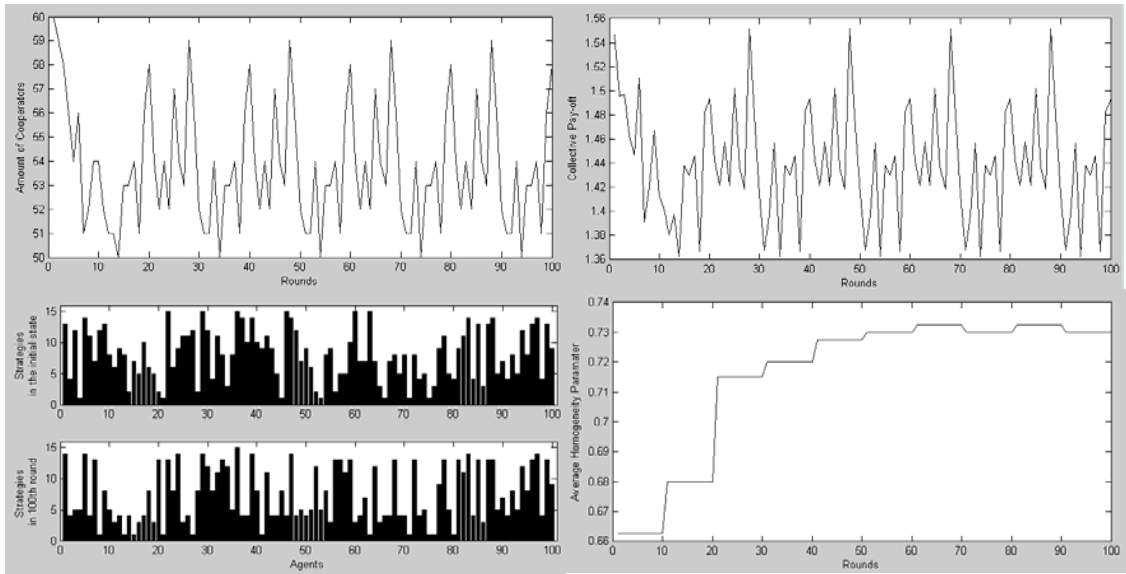

**Figure 8**
Experimental result with K=2 and N=100 with cyclic result at the equilibrium

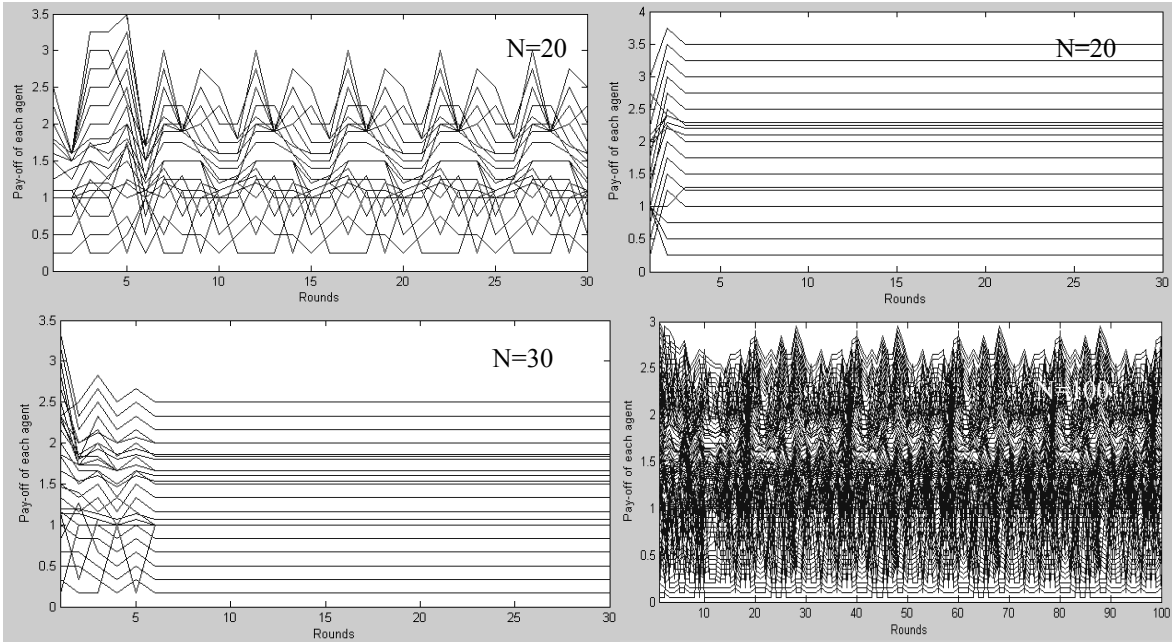

**Figure 9**
Pay-off for each agent for different number of agents

## 7. Discussion

We have described the behavior of the system whose bounded agent defect and cooperate with her neighbor agent. Every agent, based on her transition rule– choose whether to cooperate/defect (C/D) with her neighbor. At certain interval of time (state cycle length is assumed), every agent evaluate the result she gained in the form of average pay-off at the interval of time and change her strategy by using the strategy of her neighbor with highest average pay-off.

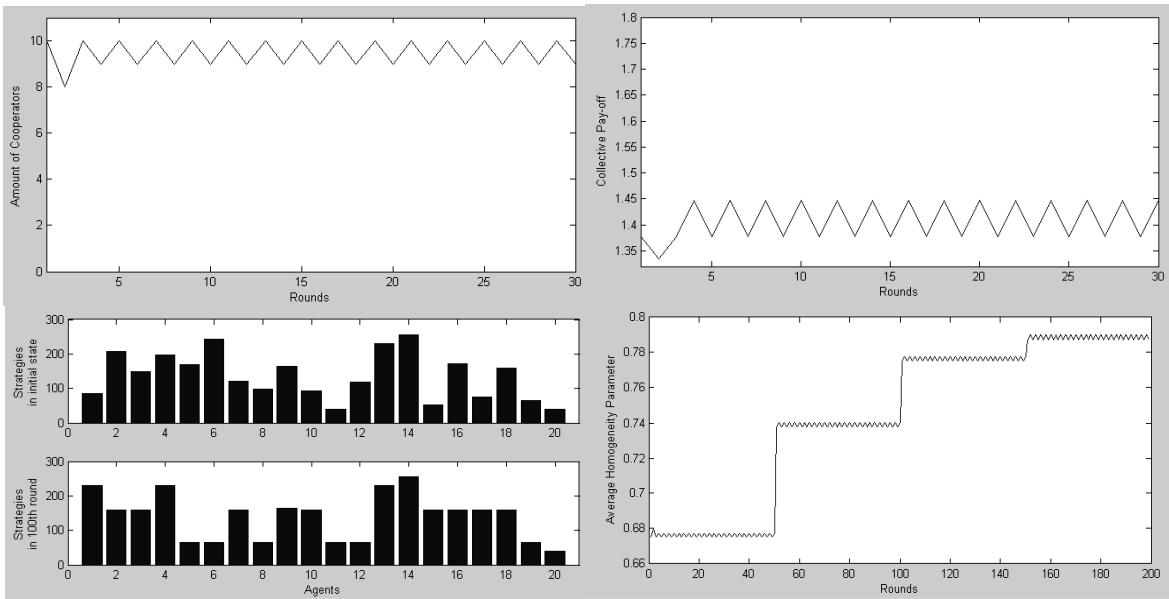

**Figure 10**
Experimental result with N=20 and K=3. The result tends to cyclic at equilibrium as the rise of homogeneity parameter

From **Figure 7** and **8**, system behavior in simulation for K=2 shows similarity with the dynamical patterns of NK Boolean Network. System will quickly move toward its steady state condition - the length of state cycle is relatively short and stable enough on perturbation (the changes of agent's strategy). This is shown by the average pay-off of the group repeating periodically in a very short interval. Pay-off value pattern for each agent is stable after the agent changes her strategy. Self-organizing patterns of the system with N=20 will be different when homogeneity variable differ. Homogeneity value is rising and tends to make system present in a frozen static regime.

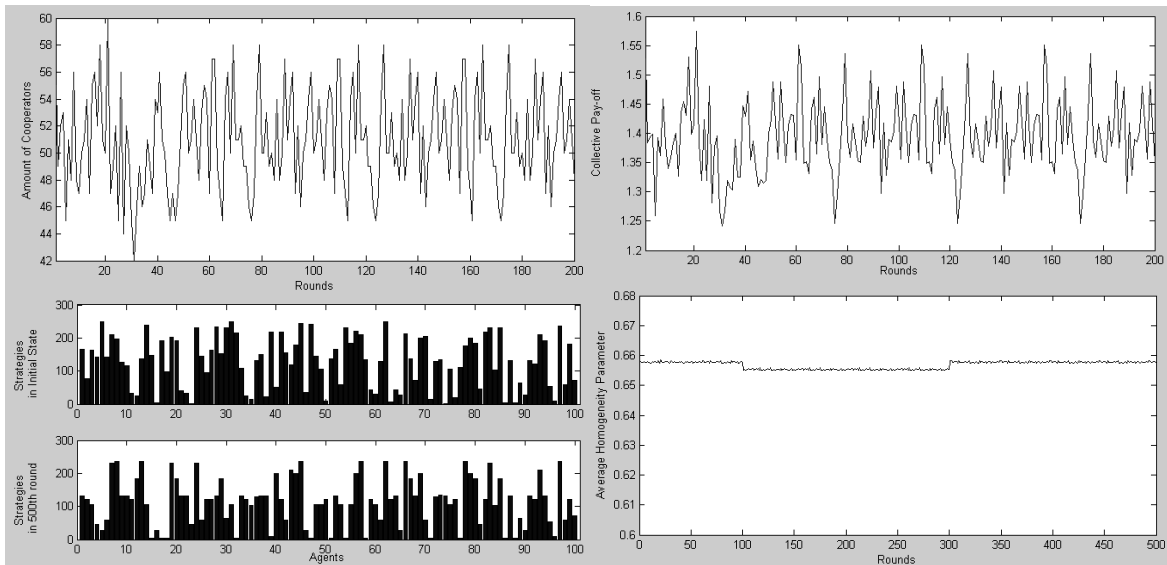

**Figure 11**
Simulation result with N=100, and K=3. State cycle length assumed less than 100, agent changes strategy at every end of state cycle

At value K=3 (**Figure 10** and **11**), system behavior is generally the same with K=2, dynamical patterns occur in the form of equilibrium appears cyclic. Yet, in the very long period, system will return to its initial condition when there is strategy change at interval 100. The increase of average homogeneity parameter explains why the system behaves differently for the same K. As shown in figures, system will be more order on the rise of its homogeneity variable.

In social system, where every person bases her action upon little information from a group of person consists of small number persons, system tends to be static, there will be no dynamical patterns and the system will be quicker to return to its equilibrium. However, if every agent bases her behavior or action on higher amount of information, then the system will tend to be relatively more dynamical and the equilibrium is assigned with in long period of time.

**Figure 9** shows that at equilibrium, player with same strategy is uncertainly will gain the same pay-off value. It is interesting since it often happens in reality where person that imitates other person's strategy does not always get the same result. Here we show that our model has performed a unique character of every agent representing social system.

In society, we can see that the increase of cooperator numbers will rise up average pay-off. Although the paper has been able to show how NIPD in RBN fashion, yet the limited computational facilities caused the impossibility to gain spectacular conclusion of the NIPD discourse– e.g.: dominant strategy like tit-for-tat model strategy in dynamic model IPD (Axelrod, 1984). Technically we can say that in this paper, we see the collective patterns caused by numbers of cooperators and pay-off value obtained by agent and her group, when she behaves as well as Boolean agent.

## 8. Conclusion

We have shown how to perform iteration over NIPD in RBN fashion. In several experiments performed we show the potential of RBN model to be utilized as alternative genetic model – as claimed by Kauffman in (Kaufman, 1993, Paczuski, *et.al*, 2000).

The smaller amount of agents whose high homogeneity will perform more stable system, a conclusion that showing that in a homogenous society conflict probability is seldom– agent tends to cooperate and system direction can be "predicted" easily. In return, social system becomes mode dynamic with the rise amount of agents. However, collective system will tend to reach its equilibrium (with certain pay-off) although it has to through a long process (large attractor length). This is the social system behavior performing the robustness in society and the fact of the self-organizing system.

The weakness of this paper is known comes from the lack computational facilities good enough by means that further works will highly depend on better facilities to gain better conclusion and theoretical exploration.

In short, we can say that RBN is an alternative to construct evolutionary model of social system with complex characters like self-organizing, adaptive, and evolve dynamically. These characters are more or less theoretically explained by RBN for organism metabolism widely – this paper is a mean of hope the same thing to view social system metabolism to be observed further.

## Acknowledgement

The writers would like to thank BFI for critics, facilitation, and climate in accomplishing the paper. We also would like to thank Yohanes Surya (*Board of* Advisory BFI) for financial support in the period in which research runs. All faults remain the auhtors'.